\magnification1200

\def\title{
{\centerline{{\bf Dynamical (super)symmetries of monopoles and
vortices}
\foot{Based on a review talk given at the {\it International Symposium on Advanced Topics in
Quantum Physics}, Shanxi'92. Ed. J.-Q. Liang, M.-L. Wang, S.-N. Qiao, D.C. Su.
pp. 283- 293. Science Press, Beijing (1993)
[Tours Preprint N. 47/92], and on Tours Preprint N. 60/93 (1993) (unpublished).}
}
}}

\def\runningtitle{
(Super)symmetries of monopoles and vortices
} 

\def\runningauthors{Horv\'athy
} 

\def\authors{
\centerline{P. A. HORV\'ATHY}\vskip 3mm

\centerline{D\'epartement 
de Math\'ematiques, Universit\'e}

\centerline{Parc de Grandmont, 
F-37200 TOURS (France)}
}

\vsize = 9.1truein 
\hsize = 6.5truein 

\headline ={
\ifnum\pageno=1\hfill
\else\ifodd\pageno\hfil\tenit\runningtitle\hfil\tenrm\folio
\else\tenrm\folio\hfil\tenit\runningauthors\hfil
\fi\fi
} 

\pageno=1
\footline = {\hfil} 


\def\parag{\hfil\break} 
\def\ccr{\cr\noalign{\medskip}}
\def\IR{{\bf R}} 
\def\IS{{\bf S}}
\def\smallcirc{{\raise 0.5pt \hbox{$\scriptstyle\circ$}}}
\def\and{\qquad\hbox{and}\qquad}
\def\where{\qquad\hbox{where}\qquad}
\def\2{{1\over 2}}
\def\D{D\llap{\big/}}
\def\bA{{\bf A}}
\def\br{{\bf r}}
\def\bq{{\bf q}}
\def\smallover#1/#2{\hbox{$\textstyle{#1\over#2}$}}
\def\2{{\smallover 1/2}}

\def\semidirectproduct{
{\ooalign
{\hfil\raise.07ex\hbox{s}\hfil\crcr\mathhexbox20D}}}
\def\ccr{\cr\noalign{\medskip}}

\font\tenb=cmmib10 
\newfam\bsfam

\textfont\bsfam=\tenb

\mathchardef\omegab="0821
\mathchardef\Gammab="0800
\mathchardef\epsilonb="080F
\mathchardef\varepsilonb="0822
\mathchardef\lambdab="0815
\mathchardef\xib="0818
\mathchardef\etab="0811
\mathchardef\pib="0819
\mathchardef\sigmab="081B


\newcount\ch 
\newcount\eq 
\newcount\foo 
\newcount\ref 

\def\chapter#1{
\parag\eq = 1\advance\ch by 1{\bf\the\ch.\enskip#1}
}

\def\equation{
\leqno(\the\ch.\the\eq)\global\advance\eq by 1
}

\def\foot#1{
\footnote{($^{\the\foo}$)}{#1}\advance\foo by 1
} 

\def\reference{
\parag [\number\ref]\ \advance\ref by 1
}

\ch = 0 
\foo = 1 
\ref = 1 

\title
\vskip 5mm
\authors

\parag
{\bf Abstract.}

{\it The dynamical (super)symmetries for various monopole systems are reviewed.  
For a Dirac monopole, no smooth Runge-Lenz vector can exist; there is, however, a spectrum-generating conformal $o(2,1)$ dynamical
symmetry that extends into $osp(1/1)$ or $osp(1/2)$ for spin
$1/2$ particles. Self-dual 't Hooft-Polyakov-type monopoles admit
an $su(2/2)$ dynamical supersymmetry algebra, which allows us to reduce
the fluctuation equation to the spin zero case. For large $r$ 
the system reduces to a Dirac monopole plus an suitable inverse-square
potential considered before by  McIntosh and Cisneros, and by Zwanziger
 in the
spin $0$ case, and to the `dyon' of D'Hoker and Vinet for spin
$1/2$. The asymptotic system admits a Kepler-type dynamical symmetry as
well as a `helicity-supersymmetry' analogous to the one Biedenharn found in
the relativistic Kepler problem. Similar results hold for the Kaluza-Klein
monopole of Gross-Perry-Sorkin. For the magnetic vortex, 
the $N=2$ supersymmetry 
of the Pauli Hamiltonian in a static magnetic field in the plane combines with the $o(2)\times o(2,1)$ bosonic symmetry into an 
$o(2)\times osp(1/2)$ dynamical superalgebra.}

\chapter{Introduction.}

The architype of a dynamical symmetry is provided by the Runge-Lenz vector [1]
in the Kepler problem,
$$
{\bf A} = \2\left\{{\bf p}\times{\bf L} -{\bf L}\times{\bf p}\right\}
-\rm{M}{\hat{\bf r}}, 
\equation
$$
where $M$ is the mass of the Sun, the planet's mass is taken to be $1$, and
${\bf L}$ denotes the planet's (orbital) angular momentum, ${\bf L} = {\bf
r}\times{\bf p}$. The vector ${\bf A}$ is directed from the Sun's position
towards the perihelion point. Under commutation [Poisson bracket] the 
Runge-Lenz vector and the angular momentum close into
$o(4)$ for bound (elliptic) motions, into $
o(3)\oplus_s\IR^3$ for parabolic motions and into $o(3,1)$ for
hyperbolic motions. This makes it possible to calculate the spectrum and the
$S$-matrix algebraically. 

The Kepler problem also admits an $o(2,1)$ \lq spectrum-generating 
symmetry', which combines with the $o(4)/o(3,1)$ into an
irreducible representation of the conformal group $o(4,2)$ [1].

In this Review we examine how similar dynamical symmetries -- as
well as supersymmetries -- arise for various magnetic
monopole systems.
In the last Chapter, we examine what happens around
a magnetic vortex.

\goodbreak
\chapter{The Dirac monopole.}  [2]

Let us consider a Dirac monopole, whose magnetic field is
$$
{\bf B} = g\,{{\bf r}\over r^3}.
\equation
$$
The conserved angular momentum of a charged, spinless particle is
$$
{\bf L}_0 = {\bf r}\times{\pib} - q{\hat{\bf r}},
\equation
$$
where $\pib = {\bf p} - iq{\bf A}_D$, rot ${\bf A}_D = {\bf r}/r^3$, $q =
eg$, $e$ being the electric charge. Since 
$$
{\bf L}_0.{\hat{\bf r}} = - q,
\equation
$$
the particle moves classically on a cone of opening angle $\cos\alpha =
-q/L_0$. There are no bound motions.

The problem of having a conserved Runge-Lenz-type vector naturaly arises, and
it has been claimed  [3] that
the vector ${\bf A}$ which points from the origin to the closest (\lq
perihelion') point of the trajectory is such a conserved vector, which
would generate, with the angular momentum, an $o(3,1)$ dynamical
symmetry. This statement is, however, false: a Dirac monopole cannot
admit any time-independent, conserved Runge-Lenz-type vector [4]. This can be
understood by considering the \lq umbrella' transformation of Boulware et al.
[5],  
$$
{\bf r}\;\mapsto\;{\bf R}\quad =\quad {{\bf r} - {\hat{\bf L}}_0
({\bf r.}{\hat{\bf L}}_0)\over\sin\alpha},
\equation
$$
which rotates the monopole problem into a potential problem: the particle
trajectories in the monopole field correspond to those in the plane
perpendicular to the angular momentum, ${\bf L}_0$, in an $-q^2/2R^2$
potential [and makes the  $o(2,1)$ symmetry [6] manifest]. 

The inverse-square potential problem is integrable. Golo's \lq
Runge-Lenz' vector goes thereby into the vector pointing to the
closest point, ${\bf R}_0$, of the rotated trajectory in  the plane
perpendicular to ${\bf L}_0$. This transformation is, however, {\it singular}
when the motion is radial: when the cone's opening angle closes to zero, the
direction of the umbrella transformation becomes undetermined. More precisely,
the inverse transformation becomes the familiar Hopf fibering $U(1)\to
SO(3)\to\IS^2$ [4].
 
 A spinless particle in the field of a Dirac monopole admits instead 
 an $o(2,1)$
symmetry [6], generated by the \lq non-relativistic conformal
transformations' 
$$
\matrix{
H \;= &\2{\pib}^2&\qquad\hbox{time translations}\ccr
D \;= &\;tH - {1\over 4}\{\pib, {\bf r}\}
&\qquad\hbox{dilations}\ccr
K \;=&-t^2H + 2tD + \2mr^2\;
&\qquad\hbox{expansions}
\cr
}
\equation
$$
which satisfy the $o(2,1)$ relations
$$
\matrix{
&[H, D]= iH, &[H,K] = 2iD, &[D,K] =iK,
\cr
}
\equation
$$
allowing for a derivation of the spectrum from the group theory. 

This result can be explained from studying the {\it non-relativistic
structure of space-time}. A free, non-relativistic particle admits in fact
the so-called  Schr\"odinger group as symmetry [7]. This
latter is the extension of the Galilei group with dilations and expansions. It
is best understood in the five-dimensional framework, where non-relativisic
motions are lightlike reductions of null geodesics in a five-dimensional
Lorentz manifold [8]. Jackiw's $o(2,1)$ is just the
residual symmetry left over from the Schr\"odinger group after adding
a Dirac monopole.  

The only potential which is consistent with the conformal algebra (2.5) is
$\lambda^2/r^2$~: for an arbitrary $\lambda$, the Hamiltonian
$$
\2\,\left[\pib^2+{\lambda^2\over r^2}
\right]
\equation
$$
is $o(2,1)$ symmetric. Adding a Coulomb term would break 
 this symmetry.
However, as first noticed by Zwanziger, and by McIntosh and Cisneros (MCZ)
[9], a slightly different system {\it does} have a Kepler-type dynamical
symmetry. It consists of Dirac monopole plus a fine-tuned inverse-square
potential plus a Coulomb term,  
$$
H_{MCZ} = \2\Big({\pib}^2 + {\alpha\over r} + {q^2\over r^2}\Big),
\equation
$$
which admits a conserved Runge-Lenz vector, namely
$$
{\bf A}_0 = \2\left\{{\pib}\times{\bf L}_0 -{\bf L}_0\times\pib\right\} 
- q^2{\hat{\bf r}}.
\equation
$$
This is understood by noting that, when applying the \lq
umbrella-transformation' (2.4) the $q^2/2r^2$ potentials cancel and we are left
with an effective Kepler problem. The $o(4)/o(3,1)$ dynamical symmetry generated
by ${\bf L}_0$ and ${\bf A}_0$ can be used to determine the spectrum and the  scattering matrix [9], respectively. It extends into $o(4,2)$, but in another
representation as for Kepler [1,10].

Jackiw's result was generalized [11] to a spin $\2$ particle with
gyromagnetic ratio $2$, described by the two-component Pauli Hamiltonian  
$$
H_P = \2\Big(\pib^2 - q\,{\sigmab.\hat{\bf r}\over r^2}\Big).
\equation
$$
This system has not only the bosonic $o(2,1)$ with $D$, $K$ in (2.5) as for
spin $0$, but also two conserved supercharges, namely
$$
Q = {1\over\sqrt{2}}\,\sigmab.\pib
\and
S = {1\over\sqrt{2}}\,\sigmab.{\bf r} -tQ,
\equation
$$
which close with the bosonic generators into an $osp(1/1)$ superalgebra, i.e.
(2.6), supplemented by
$$
\matrix{
&[Q,D]={i\over 2}Q, &[K,D]=iS, &[S,H] = iQ, &[S,D]=-{i\over 2}S,
\ccr
&[Q,H]=0, &[S,K]=0, & &
\ccr
&\{Q,Q\}=2H &\{Q,S\}=-2D, &\{S,S\}=2K. &\cr
}
\equation
$$
The $osp(1,1)$ symmetry, which allows to derive the spectrum algebraically, can
be seen to be the residual superalgebra of the \lq super-Schr\"odinger
algebra', obtained from adding the (fermionic) \lq helicity operator' $Q$ in  (2.11) to
the Schr\"odinger group [12].

\chapter{Supersymmetric quantum mechanics.}

D'Hoker and Vinet [13,14] have further generalized the problem. To
explain their results, let us consider a four-dimensional, euclidean space and
choose the representation 
$$
\gamma^k=\pmatrix{0&\sigma^k\ccr \sigma^k&0\cr},
\qquad
\gamma^4=\pmatrix{0&-i1_2\ccr i1_2&0\cr},
\qquad
\gamma^5=\pmatrix{1_2&0\ccr0&-1_2\cr},
\equation
$$
for the Dirac matrices. Let $A_{\mu}$ denote a gauge field. The
four dimensional Dirac operator,
$$
\D\equiv \gamma^{\mu}(\partial_\mu-iA_{\mu}) \equiv\pmatrix{&Q^{\dagger}\cr Q&\cr}
\equation
$$
is, as in any even dimensions {\it chiral-supersymmetric}. This means that the
square of $\D$, 
$$ 
\D^2 = \pmatrix{H_1&\cr&H_0\cr}, 
\equation
$$
is a supersymmetric hamiltonian. Its
$\pm 1$ chirality sectors (eigensectors of $\gamma^5$) 
are related by the unitary transformations 
$$
U=Q\,{1\over\sqrt{H_1}}
\and  
U^{-1}\equiv U^{\dagger} = 
{1\over\sqrt{H_1}}\,Q^{\dagger},
\equation
$$
which intertwine $H_1=Q^{\dagger}Q$ and $H_0=QQ^{\dagger}$, 
$H_1 = U^{\dagger}H_0U$. If $\Psi_0$ is an
$H_0$-eigenfunction with eigenvalue $E > 0$, then
$$
\pmatrix{
U^{\dagger}\Psi_0\cr\pm\Psi_0\cr
}
\equation
$$
is a $\D$-eigenfunction with eigenvalues $\pm 
\sqrt {E}$. Zero-energy ground states may arise; the difference of their
multiplicities in the two sectors, called the Atiyah-Singer index, is
calculated by topological formulae. 

Furthermore, if $A_0$ is conserved for $H_0$, $[A_0,H_0] = 0$, then
$$
A_1=U^{\dagger}A_0U
\equation
$$
is conserved for $H_1$, $[A_1,H_1] = 0$.

Let us first apply these framework to the gauge field
$$
{\bf A} = q{\bf A}_D,
\qquad
A_4 = {\lambda\over r},
\equation
$$
where $\lambda$ is an arbitrary real constant. This gauge field represents
a Dirac monopole plus a Coulomb potential in the fourth
(euclidean) direction. Assuming that nothing depends on the fourth
direction, $\partial_4(\ .\ )=0$, the associated Dirac operator becomes
$$\eqalign{
{1\over\sqrt{2}}\D = &{1\over\sqrt{2}}\pmatrix{&Q^{\dagger}\ccr Q&\cr} =
{1\over\sqrt{2}}\Big(\gamma^i\pi_i + \gamma^4{\lambda\over
r}\Big) =\ccr
&{1\over\sqrt{2}}\ \pmatrix{ &\sigmab.\pib 
-i{\displaystyle\lambda\over\displaystyle r}\ccr
\sigmab.\pib+i{\displaystyle\lambda\over\displaystyle r}&\cr}.\cr
}
\equation
$$
Its square is the four-component
Hamiltonian   
$$
H = \2\,\left\{\pib^2-q{\sigmab.\hat{\bf r}\over r^2}+{\lambda^2\over r^2} 
-\lambda\gamma^5 {\sigmab.\hat{\bf r}\over r^2}\right\}.
\equation
$$
The Hamiltonian (3.9) is block-diagonal, and
the $\pm 1$ chirality components only differ in the sign of $\lambda$. They  
describe two uncoupled spin $\2$ particles with anomalous gyromagnetic ratios.

Interestingly, the Hamiltonian (3.9) is a perfect square in {\it two different
ways}~:  
$$ 
Q_1 = {1\over\sqrt{2}}\gamma^5\Big(\gamma^i\pi_i +
\gamma^4{\lambda\over r}\Big)
\and
Q_2 = -i\gamma^5Q_1
\equation
$$
both satisfy $\{Q_a,Q_b\} = \delta_{ab}H$, and are hence conserved. They mix
with the bosonic $o(2,1)$ symmetry, yielding two more supercharges, namely
$$
S_1 =-tQ + {1\over\sqrt{2}}\ \gamma^5\ \gamma^ir_i
\and
S_2 = -i\ \gamma^5S_1
$$
which satisfy $\{S_a,S_b\} = 2\delta_{ab}K$. Finally,
$$
\{Q_a, S_b\} = -2\delta_{ab}D + 2\epsilon_{ab}Y,
\equation
$$
where $Y$ is the parity operator
$$
Y =\2\ \gamma^5\Big(\sigmab.\ell+ {3\over 2} -
\lambda\ \gamma^5{\sigmab.\hat{\bf r}\over r}\Big), 
\where 
\ell = {\bf r}\times\pib.
\equation
$$
The four bosonic operators $H, D, K, Y$ close with the fermionic operators
$Q_a, S_a$ (a = 1, 2) into the superalgebra $osp(1/2)$.

Since the field (3.7) is manifestly spherically symmetric, 
the total angular momentum, 
$$
{\bf J} = {\bf L}_0 + \2\ \sigmab
\equation
$$
is also conserved. 

For the special value $q =\pm\lambda$,
the Pauli term drops out from one of the sectors while the gyromagnetic ratio
becomes $4$ in the other. Eq. (3.9) reduces hence to
$$
H = \pmatrix{H_1&\ccr &H_0\cr} = 
\2\,\left\{\pib^2+{\lambda^2\over r^2} +{2q\over r^2} 
\pmatrix{\sigmab.\hat{\bf r}
\ccr
0\cr}
\right\}. \equation
$$
Being spin-independent,
the lower Hamiltonian clearly admits 
$$
{\bf S}_0 = \2\sigmab
\equation
$$
as symmetry. But supersymmetry implies that its partner Hamiltonian has also
a \lq spin' symmetry,
$$
{\bf S}_1= U^{\dagger}{\bf S}_0U
\equation
$$
commutes with $H_1$. ${\bf S}_0$ and ${\bf L}_0 = {\bf J} - {\bf S}_0$ are
hence both conserved for $H_0$. Thus, ${\bf S}_1$ and  
$$
{\bf L}_1=U^{\dagger}{\bf L}_0U={\bf J}-{\bf S}_1 = 
{\bf L}_0 + \2\ \sigmab -{\bf S}_1
\equation
$$
are both conserved for $H_1$
The combined system $\pmatrix{H_1 &\cr &H_0\cr}$  has therefore two
conserved \lq angular momenta', namely 
$$
{\bf S}= \pmatrix{{\bf S}_1&\ccr &{\bf S}_0\cr}
\and
{\bf L}= \pmatrix{{\bf L}_1&\ccr &{\bf L}_0\cr}.
\equation
$$

The action of the supercharges extends the $o(3)_{spin}$
algebra into $u(2/2)$. Let us indeed define the vector supercharges
$$
{\bf Q}_{\alpha} = 2i[{\bf S}_0, Q_{\alpha}]
\qquad\qquad(\alpha = 1,2),
\equation
$$
i.e.
$$
{\bf Q}_1 = \pmatrix{ &-2iQ^{\dagger}{\bf S}_0\ccr 2i{\bf S}_0Q &\cr},
\qquad
{\bf Q}_2 = \pmatrix{ &-2Q^{\dagger}{\bf S}_0\ccr -2{\bf S}_0Q &\cr}.
\equation
$$
All these operators commute with the Hamiltonian $H$. One has furthermore
$$
[\gamma^5, bosonic] = 0
\qquad
\big\{\gamma^5, fermionic\big\} = 0.
$$
To summarize, the bosonic operators ${\bf S}_0,{\bf S}_1,\gamma^5, H$ and the
fermionic  operators ${\bf Q}_a, Q_a$ satisfy the
(anti)\-commu\-tation relations
$$
\left.\eqalign{
&[S_0^i, S_0^j] = i\epsilon_{ijk}\,S_0^k,\ccr
&[S_1^i, S_1^j] = i\epsilon_{ijk}\,S_1^k,\hskip 45mm\ccr
&[S_0^i, S_1^j] = 0\ccr
&[\gamma_5, S_0^i] \;=\;[\gamma_5,S_1^i]\;=\;0\cr}\right\}\qquad
\hbox{bosonic sector}
$$
$$\left.\eqalign{
&[\gamma^5, Q_a] =
2i\epsilon_{ab}Q_b\qquad (a,\ b = 1,2)\ccr
&[\gamma^5, Q_a^k] =
2i\epsilon_{ab}Q_b^k\qquad (a,\ b = 1,2,\;k = 1,2,3)\ccr
&[S_0^i,Q_a^j] = {i\over 2}\,\big(\delta_{ij}Q_a +
\epsilon_{ijk}Q_a^k\big),\ccr
&[S_0^i, Q_a^j] = -{i\over 2}\,Q_a^i,\ccr
&[S_1^i,Q_a^j] =  -{i\over 2}\,\big(\delta_{ij}Q_a -
\epsilon_{ijk}Q_a^k\big),\ccr
&[S_1^i, Q_a] = {i\over 2}\,Q_a^i
\cr}\right\}
\qquad\matrix{\hbox{action of}\ccr\hbox{bosonic operators}\ccr
\hbox{on fermionic sector}\ccr}
$$
$$
\equation
$$
$$
\left.\eqalign{
&\{Q_a,Q_a\} = 2\delta_{ab}H,\ccr
&\{Q_a,Q_b^i\} = -4H\epsilon_{ab}(S_0^i + S_1^i),\ccr
&\{Q_a^i,Q_b^j\} = 2H\delta_{ij}\delta_{ab} -
4H\epsilon_{ijk}\epsilon_{ab}(S_1^i - S_1^i),\hskip 5mm\ccr
}\right\}
\qquad\hbox{fermionic sector}
$$
i.e. close into the $su(2/2)$ SUSY algebra [14, 17, 18]. The $osp(2,1)$
found before mixes with the $o(3)_{rotations}$ and the $u(2/2)_{spin}$ to yield
a supersymmetric version of $o(4,2)$. Its precise structure has not yet been
determined.

\chapter{Self-dual 't Hooft-Polyakov monopoles.}

The Dirac monopole was generalized by 't Hooft and Polyakov in non-Abelian
gauge theory [2]. It is a static, purely magnetic ($\partial_0 = 0$), 
everywhere-regular, finite-energy solution to the $SU(2)$ Yang-Mills Higgs
equations associated to the energy functional
$$
E = \int d^3x\ \left\{{1\over 4}\,Tr(F_{ij}F^{ij}) + \2\,Tr(D_j\Phi D^j\Phi)
+ {\lambda\over 4}\big[1 -Tr(\Phi^2)\big]^2\right\},
\equation
$$
where $F_{ij} = \partial_iA_j - \partial_jA_i + [A_i,A_j]$ and $D_j\Phi =
\partial_j\Phi + [A_j,\Phi]$.

Finite-energy requires $\vert\Phi\vert\simeq 1$ for large $r$, so that the
asymptotic values of the Higgs field define a mapping from the \lq sphere
at infinity' $\IS^2$ into the \lq vacuum manifold' ${\cal M}=\vert\Phi\vert =
1$. ${\cal M}$ is again a two-sphere, so it provides us with the integer  
$$
n=[\Phi]\in\pi_2(\IS^2)\simeq{\bf Z},
\equation
$$
called the topological charge. 

For non-vanishing Higgs potential (i.e. $\lambda\neq 0$), the sytem has the
same $o(2,1)$ bosonic symmetry as the Dirac monopole.

In the \lq
Prasad-Sommerfield limit' of vanishing $\lambda$ the situation is
different. The second-order field equations associated to (4.1) are solved by
the \lq self-duality' or \lq Bogomolny' equations 
$$
{\bf B} = {\bf D}\Phi
\where
B_i = \2\epsilon_{ijk}F^{jk}.
\equation
$$
For $n=1$ for exemple, Prasad and Somerfield found the solution
$$
A_j^a = \epsilon_{ajk}\Big(1 - {r\over\sinh r}\Big)\,{x^k\over r^2},
\qquad\qquad
\Phi^a = -(\coth r - {1\over r})\,{x^a\over r}.
\equation
$$
Setting $A_4 = \Phi$ and requiring $\partial_4 = 0$, a PS monopole can also be
viewed as a self-dual Yang-Mills field in four euclidean dimensions.

Let us now consider a massless Dirac particle in a BPS background, described
by the four dimensional Dirac operator
$$
\D =\pmatrix{&Q^{\dagger}\cr Q&\cr} =
\pmatrix{ &{\sigmab.\pib}-i\Phi\cr{\bf\sigmab.\pib}+i\Phi\cr}.
\equation
$$
As explained in Section 3, $\D$ is chiral-supersymmetric. 
Now, owing to
$$
QQ^{\dagger}=\pib^2+\Phi^2 +\sigmab.({\bf B}-{\bf D}\Phi),
\qquad
Q^{\dagger}Q=\pib^2+\Phi^2 +\sigmab.({\bf B}+{\bf D}\Phi),
$$
the spin drops out in the self-dual sector, while we get a factor $2$ in
the other one: $H_0$ describes two {\it spin $0$} particle (or a spin $\2$
particle with gyromagnetic ratio 0), while $H_1$ describes a particle with {\it
anomalous gyromagnetic ratio $4$}. This is why the fluctuation equation  in the
BPS background can be reduced to the study of the spin $0$ system [15, 16]. 
	
The spin operator ${\bf S}_0 =\sigmab/2$ is trivially conserved for $H_0$.
Its superpartner,
$$
{\bf S}_1=U^{\dagger}{\bf S}_0U ={1\over H_1}\,
\left\{\2\big[\pib^2 - \Phi^2\big]\sigmab +
\Phi\,(\pib\times\sigmab)-(\sigmab.\pib)\,\pib\right\},  
\equation
$$
is therefore conserved for $H_1$. 

Zero-energy ground states only arise for $H_1$ (but not for $H_0$) as solutions
of  $Q\Psi = 0$. The multiplicity of these states (the Atiyah-Singer
index) was found to be $2n$, twice the topological charge [16].

Since BPS monopoles with topological charge $n \geq 2$ are not spherically
symmetric, for a general BPS monopole this is the end of the story. For the
$n=1$ the BPS solution above, however, we also have spherical symmetry and hence
the total angular momentum, 
$
{\bf J} = {\bf L}_0 + \2\sigmab,
$ is conserved. The same argument as in Section 3 shows that
$$
{\bf L}_0 = {\bf J}-{\bf S}_0 
\and
{\bf L}_1=U{\bf L}_0U^{\dagger}={\bf J}-{\bf S}_1
$$
 cf. (3.17)
are conserved for $H_0$ and $H_1$ respectively; the commuting operators
$
{\bf L}$
and 
${\bf S}$
in Eq. (3.18) generate
$
o(3)_{rotations}\oplus o(3)_{spin},
$
and the spin part is extended into $u(2/2)$ as in Eq. (3.21).

\goodbreak

\chapter{Dyons.}

For large $r$, the systems become even more symmetric. The
BPS monopole becomes an imbedded Dirac monopole with an additional long-range
scalar field $\Phi\sim 1 - 1/r$. For eigenstates of the electric
charge operator $Q_{em} = {\hat\Phi}$ the $SU(2)$-covariant derivative reduces
to the electromagnetic covariant derivative with coupling constant equal $q$,
the electric charge. Thus,
$$
\eqalign{
&H_0\;\longrightarrow\;
H_{MCZ}=\pib^2 + q^2\Big(1 - {1\over r}\Big)^2
\ccr
&H_1\;\longrightarrow\;H_D=\pib^2 + q^2\Big(1 - {1\over r}\Big)^2 +
2q{\sigmab {\bf .r}\over r^3}\quad
\cr}
\qquad\hbox{when}\quad r\to\infty.
\equation
$$

Remarkably, the large-$r$ limit of $H_0$ is precisely the $H_{MCZ}$, the 
McIntosh - Cisneros - Zwanziger Hamiltonian (2.8) (times
the unit $2\times 2$ matrix), while its partner $H_1$ becomes
the \lq dyon' Hamiltonian $H_D$ of D'Hoker and Vinet [17,18]. Supersymmetry
then converts the Runge-Lenz vector ${\bf A}_0$ of MCZ into a spin-dependent
Runge-Lenz vector,  
$$
{\bf A}_1 =  \underbrace{\left\{{1\over 2}\Big(\pib\times{\bf L}_0 -
{\bf L}_0\times\pib\Big) - q^2 {\bf\hat r}\right\}}_{{\bf A}_0} + \,
\pib\times\sigmab + {q\over r}\sigmab -
q {{\bf r.\sigmab}\over r^3}\,{\bf r} - {q\over 2}\,\sigmab, 
\equation
$$
which is conserved for $H_D$. For the asymptotic system
$$
\pmatrix{
H_D&\ccr &H_{MCZ}\cr
},
\equation
$$
the bosonic symmetry algebra $o(3)_{rotations}\oplus o(3)_{spin}$ extends
therefore into 
$$
o(4)\oplus o(3)_{spin}
\equation
$$
for bound motions (and into $o(4)\oplus o(3)_{spin}/  
o(4)\oplus o(3)_{spin}$ for scattered motions)\foot{It is likely that this
symmetry is further extended to $o(4,2)\oplus o(3)_{spin}$.}, generated by 
$$
{\bf A} = \pmatrix{
{\bf A}_1 &\ccr &{\bf A}_0\cr
},
\equation
$$
and by ${\bf L}$ and ${\bf S}$ in Eq. (3.17), to which is added the
supersymmetry algebra $u(2/2)$ in Eq. (3.21).

The dynamical symmetry (5.4) makes it possible to find the spectrum [14,18,19], 
$$
E = q^2\bigg(1-{q^2\over p^2}\bigg),
\qquad
p=\matrix{q,\,q+1,\ldots\ccr q+1,\ldots\cr}
\qquad\hbox{for}\qquad
\matrix{H_1\ccr H_0\cr}.
\equation
$$
Chiral SUSY means that the spectra of $H_0$ and of $H_1$ are identical up to
zero-energy ground-states. Closer inspection shows, however, even more
symmetry, namely a {\it two-fold degeneracy}. 

Let us focus our attention to a fixed $j=const$. sector. The pattern is
reminiscent of a supersymmetric system except that the ground state energy is
nonzero. 

Generalizing Biedenharn's approach to the relativistic
Kepler problem [20], we can exhibit another conserved operator, namely
$$
{\cal R} =\pmatrix{&R^{\dagger}\cr R&\cr} =\pmatrix{
&i{\bf \sigmab.\pib} + {\displaystyle q\over\displaystyle r} +
({\bf\sigmab}.{\hat{\bf r}}){\displaystyle q^2\over\displaystyle x}
\ccr
i{\bf\sigmab.\pib} - {\displaystyle q\over\displaystyle r} + ({\bf\sigmab}. 
{\hat{\bf r}}) {\displaystyle q^2\over\displaystyle y} &\cr
}
\equation
$$
we call \lq dyon helicity' operator [19]. Here
$$
\eqalign{&x = \sigmab.\ell + 1-q{\hat\sigmab}.{\hat{\bf r}}\ccr
&y = \sigmab.\ell +1+q{\hat\sigmab}.{\hat{\bf r}}\cr}
\qquad\hbox{is conserved for}\qquad
\eqalign{&H_0\ccr&H_1\cr}\quad (\ell = {\bf r}\times{\bf p}).
\equation
$$
$x$ and $y$ both have the eigenvalues $\pm(j+1/2)$ [18,19]. They are just
the components of the Biedenharn-Temple operator
$$
\Gamma = - (\sigmab.\ell + q\ \gamma^5\ \sigmab.{\hat r}) = 
\pmatrix{-y&\ccr &-x\cr}.
\equation 
$$
Since the dyon helicity operator ${\cal R}$ satisfies
$$
{\cal R}^2 = \D^2 - E_0^{(j)}.
\equation
$$
Subtracting the ground-state energy $E_0^{(j)}$, 
$$
\D^2 - E_0^{(j)} = \pmatrix{H_1- q^2 +\displaystyle{q^4\over (j+1/2)^2}&\cr
&H_0- q^2 +\displaystyle{q^4\over (j+1/2)^2}\cr} 
\equation
$$
becomes hence a supersymmetric, with ${\cal R}$ as square-root.
The new supersymmetry-sectors are the $\pm 1$ eigensectors of the
normalized Biedenharn-Temple operator $\Gamma/\vert\Gamma\vert$. 

The dyon helicity operator has the nice property that it respects
the angular decomposition. Explicit  eigenfunctions are constructed in 
Ref. [19].

\chapter{Particle in the Wu-Yang monopole field.}

The MCZ system has yet another
symmetric generalizations. Rather then considering spin $\2$ 
particles, with vanishing isospin,
 we can also study spin $0$ particles with {\it isospin},
 moving in a
self-dual Wu-Yang [21] monopole field. This latter is obtained by imbedding the
Dirac monopole into $SU(2)$ gauge theory and adding a suitable \lq hedgehog'
scalar field,  
$$
{\bf A} = {i\over 2r}\,\sigmab\times\hat{\bf r},
\qquad
\Phi={i\over 2}\,\Big(1-{1\over r}\Big)\,\sigmab.\hat{\bf r}.
\equation
$$

The electric charge is defined [2] as the eigenvalue of
$$
Q_{em} =\hat\Phi =\sigmab.\hat{\bf r}.
\equation
$$
The  Hamiltonian is hence
$$
H_{WY}=\2\,\big(-i\nabla-Q_{em}{\bf A}_D\big)^2+{Q_{em}^2\over2}
\Big(1-{1\over r}\Big)^2.
\equation
$$
Since on the $Q_{em} =\pm q$, eigensectors $H_{WY}$ reduces to the MCZ
Hamiltonian, such a particle admits the conserved Runge-Lenz vector [22]
$$
\eqalign{
{\bf A} = \;&\2\,\big\{
\pib\times{\bf J}-{\bf J}\times\pib\big\}-q^2\hat{\bf r}\ccr
&+{1\over 2r}\,
\left\{
q\sigmab - q(\sigmab.\hat{\bf r})\hat{\bf r} - 
-{i\over 2}\sigmab\times\hat{\bf r} -
r\sigmab\times\pib 
+(\sigmab.\ell+1)\hat{\bf r}
\right\}. 
\cr
}
\equation
$$

A variation of the model can be obtained by considering  \lq nucleon-type'
particles [13, 23],  whose electric charge operator is
$$
Q_{em}' = Q_{em}-\2\sigmab.\hat{\bf r}.
\equation
$$
The associated  Hamiltonian is only slightly different from 
yet another one
studied by D'Hoker and Vinet [13], namely 
$$
H_N =\2\bigg\{\Big(-i\nabla -(Q_{em}-\2\sigmab.\hat{\bf r}){\bf A}_D\Big)^2 
+{q^2+1/4-\sigmab.\hat{\bf r}/2\over r^2}+{\alpha\over r}\bigg\}.
\equation
$$
This  admits again a conserved Runge-Lenz vector, namely [23]
$$
\eqalign{
{\bf A} = \;&\2\,\big\{
\pib\times{\bf L}_0 - {\bf L}_0\times\pib\big\} - q^2\hat{\bf r}\ccr
&+{1\over 2r}\,\left\{q\sigmab - q(\sigmab.\hat{\bf r})\hat{\bf r} - 
-{i\over 2}\sigmab\times\hat{\bf r} -
r\sigmab\times\pib 
+(\sigmab.\ell+1)\hat{\bf r}\right\}. 
\cr
}
\equation
$$
D'Hoker and Vinet have also proved that $H_N$ is actually a
partner Hamiltonian of a supersymmetric system, namely of
$$
\pmatrix{H_N &\cr &H_D\cr}.
\equation
$$

\chapter{The Kaluza-Klein monopole.}

The Kaluza-Klein monopole [24] is obtained by imbedding the Taub-NUT
gravitational instanton as a static soliton in Kaluza-Klein theory. This
latter is described by the 4-metric  
$$
\eqalign{
&V\Big\{dr^2 + r^2 (d\theta ^2 + \sin^2\theta d\phi^2)\Big\} +
{1\over V}\Big\{ d\psi + 4m\cos\theta d\phi\Big\}^2\cr
&\hbox{where}\quad
V = 1 + {4m\over r}\ .\cr}
\equation
$$
The \lq vertical' variable $\psi$ describes a internal
circle. The apparent singularity at $r=0$ is unphysical if  $\psi$ is periodic
with period $16 \pi m$. In the usual context, the Taub-NUT
parameter, $m$ is positive. We shall, however, also consider  
$m<0$.
Such a situation arises, e.g. in the long-range scattering of self-dual $SU(2)$
monopoles [25].

$\partial_{\psi}$ is a Killing vector, and the associated conserved quantity,
$q$ is quantized in half-integers. It is identified with the
electric charge.

The curved-space gamma - matrices ${\hat\gamma}^A$ and the spin connection
$\Gamma^A$ in the KK monopole background are found to be
$$
{\hat\gamma}^j =\pmatrix{0 &-
{i\over\sqrt{V}}\,\sigmab \cr{i\over\sqrt{V}}\,\sigmab &0\cr}
\qquad
{\hat\gamma}^4 = \pmatrix{0 &\sqrt{V} + {i\over\sqrt{V}}\,\sigmab {\bf .A}
\cr\sqrt{V}-{i\over\sqrt{V}}\,\sigmab {\bf .A} &0 \cr}
\equation
$$
and
$$
\Gamma^i = \pmatrix{-{1\over\displaystyle 2V^2}(\sigmab {\bf .B}) A^i 
+ {1\over\displaystyle 2V}
\,({\bf B}\times\sigmab)^i &0\cr\noalign{\medskip} 0 &0\cr},
\qquad
\Gamma_4 = \pmatrix{-{1\over\displaystyle 2V^2}\,{\bf B .}\sigmab
&0\cr\noalign{\medskip} 0 &0\cr}. 
\equation
$$

Requiring that all fields be equivariant with respect to the
vertical action $\psi\mapsto\psi +\alpha$ i.e. have the form,
$
e^{iq\psi}\Psi,
$
The Dirac operator becomes [26]
$$
\D = \pmatrix{0 &Q^{\dagger}\cr Q &0\cr} = 
\pmatrix{0 &{1 \over\displaystyle \sqrt{V}}\ \sigmab{\bf .} \pib - i
{q \over 4m}\sqrt{V}\cr &\cr {1\over\displaystyle V}\ \sigmab {\bf .}
\pib \sqrt{V} + i{q\over 4m}\sqrt{V} &0 \cr}, 
\equation
$$
where
$\pib = -i{\bf\nabla} - (q/4m){\bf A}$, ${\bf A}$ being the
vectorpotential of a Dirac monopole of unit strength. (It is easy to check
that $Q$ and $Q^{\dagger}$ are each other's adjoint with respect to the
Taub-NUT volume element $V d^4x$, as they have to be).
Using the self-duality property 
$$
\nabla V  = {\bf B},
\equation
$$
the square of $\D$ is readily found to be
$$
\pmatrix{
H_0 + {1\over\displaystyle V}\ \Big[- {q\over r^2V}\sigmab . 
\hat{\bf r} + 4m{\displaystyle\sigmab 
{\bf .L}_0\over r^3 V} + {12m^2\over\displaystyle r^4 V^2}\Big]&\ccr
&{1\over\displaystyle V}\Big[\pib^2 + ({q\over 4m})^2 V^2\Big]\cr},
\equation
$$
${\bf L}_0$ being the spin-$0$ \lq monopole' orbital angular momentum,
${\bf L}_0 = {\bf r}\times\pib - q\hat{\bf r}$ in Eq. (2.2). 
(${\bf L}_0$ is conserved only for $H_0$ but not for $H_1$).
The partner hamiltonians $H_1$ and $H_0$ differ hence in a complicated
expression, and it is not at all obvious that they will have the same spectra.
Chiral SUSY implies however that this is nevertheless true.
  
Let us first focus our attention to the $\gamma^5 = -1$
sector. Observe now that the
 spin dependence has again dropped out, so it actually describes two, uncoupled,
spin $0$ particles. $H_0$ is in fact the {\it same} as
the Hamiltonian for a spin 0 particle in the KK field [4] (times the unit
matrix). 

Because the spin is uncoupled, the system again has two angular momenta,
namely orbital angular momenta and the spin vectors,
$${\bf L}_0\qquad
{\bf L}_1 = U^{\dagger}{\bf L}_0U,
\and
{\bf S}_0 = {\sigmab\over 2}
\qquad
{\bf S}_1 = U^{\dagger}{\bf S}_0U
\equation
$$
cf. (3.17-18).
$H_0$ admits [25] a Runge-Lenz vector,
$$
{\bf A}_0 = \2\{\pib \times {\bf L} - {\bf L}\times\pib\} 
- 4m\hat{\bf r}\big(H_0 - ({q\over 4m})^2\big).
\equation
$$
The vector operators ${\bf L_0}$ and ${\bf K}_0$ generate an $o(3,1)$
dynamical symmetry for scattered motions and $o(4)$ for bound
motions. Its superpartner,
$
{\bf A}_1 = U^{\dagger}{\bf A}_0U
$
[cf. (5.5)], generates an analogous dynamical symmetry group for $H_1$ [26].

\chapter{Supersymmetry of the magnetic vortex}

The three-dimensional (super)symmetries studied above 
become even larger in the plane [27,28], namely for a magnetic vortex 
(an idealization for the Aharonov-Bohm experiment).
Firstly, the $o(2,1)$ symmetry (2.5) is still present;
on the other hand, the
$N=2$ supersymmetry of the Pauli Hamiltonian of a
spin-$\2$ particle, present for any magnetic field in the plane [29],
combines, for a magnetic vortex, with
Jackiw's $o(2)\times o(2,1)$ into an $o(2)\times osp(1/2)$
superalgebra\foot{This  is to be compared with the Galilean supersymmetry
 [30] for non-relativistic
Chern-Simons systems, and with the $osp(1/2)$ found by Hughes et al. 
in a constant magnetic field [31].}.
 This curious supersymmetry
 is  realized with two (rather then four)-component 
objects, and  is only possible in two 
spatial dimensions [30]. It arises owing to the existence of
two ``scalar products" in the plane, namely the ordinary  (symmetric)
scalar product, and the (antisymmetric) vector product\foot{
The vector or cross product of two planar vectors,
${\bf u}\times{\bf v}=\epsilon_{ij}u^iv^j$, is a scalar.}.

In detail, let us first consider a spin-$\2$ particle in an
arbitrary  static magnetic field
${\bf B}=\big(0,0,B\big)$, $B=B(x,y)$.
Dropping the irrelevant $z$ variable, we  work in the plane.
Then our model is described by the Pauli Hamiltonian
$$
H={1\over2m}\left[\pib^2-eB\sigma_3\right],
\equation
$$
where $B={\rm rot}\,\bA (\equiv\!\epsilon^{ij}\partial_iA_j)$.
It is now easy to see that the Hamiltonian (8.1) is a perfect
square in {\it two different ways}~:
both operators
$$
Q={1\over\sqrt{2m}}\,\pib\cdot\sigmab
\and
Q^*={1\over\sqrt{2m}}\,\pib\times\sigmab,
\equation
$$
where $\sigmab=(\sigma_1,\sigma_2)$, satisfy
$$
\{Q,Q\}=\{Q^\star,Q^\star\}=2H.
\equation
$$
Thus, for any static, purely magnetic field in the plane, $H$ is an
$N=2$ supersymmetric Hamiltonian. The supercharge $Q$ is a standard object
 used in supersymmetric quantum mechanics;  the `twisted' charge $Q^\star$
was used, e.g., [32], to describe the Landau states
in a constant magnetic field [29,31]. 

Let us assume henceforth that $B$ is the field of a point-like
magnetic vortex directed along the $z$-axis, $B=\Phi\,\delta(\br)$,
where $\Phi$ is the total magnetic flux\foot{Our setup
can be thought of as an idealization of the spinning version 
of the Aharonov-Bohm experiment [33].}.
Inserting 
$A_i(\br)=-(\Phi/2\pi)\,\epsilon_{ij}\,\br^j/r^2$ into the Pauli
Hamiltonian $H$ in (8.1), it is  straightforward to check
that
$$
D=tH-\smallover1/4 \left\{\pib,\br\right\}
\and
K=-t^2H+2tD+\2mr^2
$$
cf. (2.3) generate, along with $H$,
an $o(2,1)$ Lie algebra  (2.6).
The angular momentum,
$J=\br\times\pib$, adds to this $o(2,1)$ an extra $o(2)$\foot{
The correct definition of angular momentum requires boundary
conditions.}.

Commuting $Q$ and $Q^\star$ with the expansion, $K$,
yields two more generators, namely
$$\eqalign{
S&=i[Q,K]
=\sqrt{m\over2}\left(\br-{\pib\over m}t\right)\cdot\sigmab,
\ccr
S^\star
&=i[Q^\star,K]
=\sqrt{m\over2}\left(\br-{\pib\over m}t\right)\times\sigmab.
\cr}
\equation
$$

It is now straightforward to see that both sets $Q,S$ and
$Q^\star,S^\star$ extend
the $o(2,1)\cong osp(1/0)$ into an $osp(1/1)$ superalgebra.
 These two algebras do not close yet, though~: the `mixed'
anticommutators $\{Q,S^\star\}$ and $\{Q^\star,S\}$ produce a new
conserved charge, {\it viz.}
$$
\{Q,S^\star\}=-\{Q^\star,S\}=J+2\Sigma,
\qquad\hbox{where}\qquad
\Sigma=\2\sigma_3.
$$
But $J$ satisfies now non-trivial
commutation relations with the supercharges,
$$[J,Q]=-iQ^\star,\quad
[J,Q^\star]=iQ,\quad
[J,S]=-iS^\star,\quad
[J,S^\star]=iS.
$$
Thus, setting
$$
Y=J+2\Sigma=\br\times\pib+\sigma_3,
$$
the generators $H,D,K,Y$ and
$Q,Q^\star,S,S^\star$ satisfy
$$
\matrix{
[Q,D]\hfill&=&\smallover i/2 Q,\hfill
&[Q^\star,D]\hfill&=&\smallover i/2 Q^\star,\hfill
\ccr
[Q,K]\hfill&=&-iS,\hfill
&[Q^\star,K]\hfill&=&-iS^\star,\hfill
\ccr
[Q,H]\hfill&=&0,\hfill
&[Q^\star,H]\hfill&=&0,\hfill
\ccr
[Q,Y]\hfill&=&-iQ^\star,\hfill
&[Q^\star,Y]\hfill&=&iQ,\hfill
\ccr
[S,D]\hfill&=&-\smallover i/2 S,\hfill
&[S^\star,D]\hfill&=&-\smallover i/2 S^\star,\hfill
\ccr
[S,K]\hfill&=&0,\hfill
&[S^\star,K]\hfill&=&0,\hfill
\ccr
[S,H]\hfill&=&iQ,\qquad\hfill
&[S^\star,H]\hfill&=&iQ^\star,\hfill
\ccr
[S,Y]\hfill&=&-iS^\star,\hfill
&[S^\star,Y]\hfill&=&iS,\hfill
\ccr
\{Q,Q\}\hfill&=&2H,\qquad\qquad\hfill
&\{Q^\star,Q^\star\}\hfill&=&2H,\hfill
\ccr
\{S,S\}\hfill&=&2K,\hfill
&\{S^\star,S^\star\}\hfill&=&2K,\hfill
\ccr
\{Q,Q^\star\}\hfill&=&0,\hfill
&\{S,S^\star\}\hfill&=&0,\hfill
\ccr
\{Q,S\}\hfill&=&-2D,\hfill
&\{Q^\star,S^\star\}\hfill&=&-2D,\hfill
\ccr
\{Q,S^\star\}\hfill&=&Y,\hfill
&\{Q^\star,S\}\hfill&=&-Y.\hfill
\cr}
\equation
$$

Added to the $o(2,1)$ relations, this means that our generators
span the $osp(1/2)$ superalgebra [11,13].
On the other hand, 
$$
Z=J+\Sigma=\br\times\pib+\2\sigma_3
$$
commutes with all generators of $osp(1/2)$, so that the full
symmetry is the direct product
$osp(1/2)\times{o}(2)$, generated by
$$
\left\{
\matrix{
Y\hfill&=&
\br\times\pib+\sigma_3,\qquad\qquad\hfill
&Q\hfill
&=&\displaystyle{1\over\sqrt{2m}}\,\pib\cdot\sigmab,\hfill
\ccr
H\hfill&=&
\displaystyle{1\over2m}\,
\left[\pib^2-eB\sigma_3\right],\qquad\quad\hfill
&Q^\star\hfill &=&
\displaystyle{1\over\sqrt{2m}}\,\pib\times\sigmab,\hfill
\ccr
D\hfill &=&
-\smallover 1/4\,
\left\{\pib,\bq\right\}
-t\displaystyle{eB\over2m}\,\sigma_3,\quad\hfill
&S\hfill
&=&\sqrt{\displaystyle{m\over2}}\,\bq\cdot\sigmab,\hfill
\ccr
K\hfill&=&
\2m\bq^2,\hfill
&S^\star\hfill&=&
\sqrt{\displaystyle{m\over2}}\,\bq\times\sigmab,\hfill
\ccr
Z\hfill&=&
\br\times\pib+\2\sigma_3,\hfill&
\cr
}\right.
\equation
$$
where we have put
$\bq=\br\-(\pib/m)t$.

The supersymmetric Hamiltonian (8.1) is the square of Jackiw's [32]
two-dimensional Dirac operator $\pib\times\sigma$.
But the Dirac operator
is supersymmetric in any even dimensional space. 
The energy levels are therefore non-negative; 
eigenstates with non-zero energy are doubly
degenerate; the system has ${\rm Ent}(e\Phi-1)$ zero-modes
[32, 33]. The superalgebra (8.6) allows for a complete
group-theoretical solution of the Pauli equation, along the lines
indicated by D'Hoker and Vinet [11,13]. 

Notice that the two-dimensional Dirac operator $\pib\times\sigma$
of Ref. [32] -- essentially our $Q^\star$ -- is
associated with the unusual choice of the two-dimensional `Dirac'
(i.e. Pauli)
matrices
$\gamma_1^\star=-\sigma_2$,
$\gamma_2^\star=\sigma_1$. 
Our helicity operator, $Q$,
is again a \lq Dirac operator' --- but one associated with the standard choice
$\gamma_1=\sigma_1$,
$\gamma_2=\sigma_2$.

\vskip 4mm
\noindent
{\bf Acknowledgements.} This review is based on joint research with 
L. Feh\'er, B. Cordani,
L. O'Raifear\-taigh, F. Bloore, C. Duval, G. Gibbons and A.
Comtet, to whom I express my indebtedness.

\bigskip


\centerline{\bf References}

\reference
B. Cordani, {\it The Kepler problem}. Birkh\"auser (2003).
The $o(4,2)$ symmetry was first found by
H. Kleinert, Colorado Lecture (1966) (unpublished);
A.O. Barut and H. Kleinert, Phys. Rev. {\bf 156} 1541, (1967);
G. Gy\"orgyi, Il Nuovo Cimento {\bf A53}, 717 (1968)

\reference
For a review on monopoles see, e.g., P. Goddard and D. Olive, 
Rep. Prog. Phys.
{\bf 41}, 1357 (1978)

\reference
Golo, JETP Lett. {\bf 35}, 535 (1982).

\reference
L. Gy. Feh\'er Journ. Math. Phys. {\bf 28}, 234 (1987).

\reference
D. G. Boulware, L. S. Brown, R. N. Cahn, S. D. Ellis, C. Lee,
Phys. Rev. {\bf D14}, 2708 (1976).

\reference
R. Jackiw, Ann. Phys. (N.Y.) {\bf 129}, 183 (1980).

\reference
R. Jackiw, Phys. Today {\bf 25}, 23 (1972);
U. Niederer, Helv. Phys. Acta {\bf 45}, 802 (1972);
C. R. Hagen, Phys. Rev. {\bf D5}, 377 (1972);
C. Duval {\it Th\`ese de Doctorat d'Etat}, Marseille (1982) (unpublished).

\reference
C. Duval, G. Burdet, H. P. K\"unzle, M. Perrin, Phys. Rev. {\bf D31}, 1841
(1985); 
C. Duval, G. W. Gibbons
and P. A. Horv\'athy, Phys. Rev. {\bf D43}, 3907 (1991). 

\reference
H. V. McIntosh and A. Cisneros, Journ. Math. Phys. {\bf 11}, 896 (1970);
D. Zwanziger, Phys. Rev. {\bf 176}, 1480 (1968); J.
Sch\"onfeld, Journ. Math. Phys. {\bf 21}, 2528 (1971); L. Gy. Feh\'er,
Journ. Phys. {\bf A19}, 1259 (1986); and in {\it Non-Perturbative
Methods in Quantum Field Theory}, Proc. 1986'Si\'ofok Conference, Horv\'ath,
Palla, Patk\'os (eds), Singapore: World Sci. (1987). For the scattering,
see L. Gy. Feh\'er and P. A. Horv\'athy, 
Mod. Phys. Lett.  {\bf A3}, 1451 (1988).

\reference
A. O. Barut and G. L. Bornzin,
Journ. Math. Phys. {\bf 4}, 141 (1971);
B. Cordani, L. G. Feh\'er and P. A. Horv\'athy,
Journ. Math. Phys. {\bf 31}, 202 (1990).

\reference
E. D'Hoker and L. Vinet, Phys. Lett. {\bf 137B}, 72, (1984).

\reference
J. P. Gauntlett, J. Gomis and P. K. Townsend,
Phys. Lett. {\bf 248B}, 288 (1990);
C. Duval and P. A. Horv\'athy,  Journ. Math. Phys. {\bf 35}, 2516 (1994)
[hep-th/0508079].

\reference
E. D'Hoker and L. Vinet, Comm. Math. Phys. {\bf 97}, 391-427 (1985).

\reference
E. D'Hoker and L. Vinet,
in {\it Field Theory, Quantum Gravity and Strings}, Meudon - Paris seminars
85/86, Springer Lecture Notes in Physics {\bf 280}, p. 156;
Lett. Math. Phys. {\bf 12}, 71 (1986).
E. D'Hoker, V. A.
Kostelecky and L. Vinet, in {\it Dynamical Groups and Spectrum Generating
Algebras}, p. 339-367, World Scientific, Singapore (1988).

\reference
E. Mottola, Phys. Lett. {\bf 79B}, 242 (1979).

\reference
E. J. Weinberg, Phys. Rev. {\bf D20}, 936 (1979). 

\reference
E. D'Hoker and L. Vinet, 
Phys. Rev. Lett. {\bf 55}, 1043 (1986).

\reference
L. Gy. Feh\'er, P. A. Horv\'athy and L. O'Raifeartaigh,
Int. Journ. Mod. Phys. {\bf A4}, 5277 (1989); 
Phys. Rev. {\bf D40}, 666 (1989).

\reference
F. Bloore and P. A. Horv\'athy,
Journ. Math. Phys. {\bf 33}, 1869 (1992).

\reference
L. C. Biedenharn, Phys. Rev. {\bf 126}, 845 (1962);
M. Berrondo and H. V. McIntosh, Journ. Math. Phys. {\bf 11}, 125 (1970).

\reference
T. T. Wu and C. N. Yang, in {\it Properties of Matter under Unusual
Conditions}, H. Mark and S. Fernbach (eds), Interscience, (1969).

\reference
A. O. Barut and G. L. Bornzin,
Phys. Rev. {\bf D7}, 3018 (1973).

\reference
P. A. Horv\'athy, Mod. Phys. Lett. {\bf A6}, 3613 (1991).

\reference
Gross and M. Perry, Nucl. Phys. {\bf B226}, 29 (1983);
R. Sorkin, Phys. Rev. Lett. {\bf 51}, 87 (1983).

\reference
N. Manton and G. W. Gibbons, Nucl. Phys. {\bf 274}, 183 (1986);
L. Gy. Feh\'er and P. A. Horv\'athy, Phys. Lett. {\bf 183B}, 182 (1987);
B. Cordani, L. Gy. Feh\'er and P. A. Horv\'athy, Phys. Lett. 
{\bf 201}, 481 (1988).

\reference
Z. F. Ezawa and A. Iwazaki,
Phys. Lett. {\bf 138B}, 81 (1984);
M.~B.~Paranjape and G.~W.~Semenoff,
   Phys.\ Rev. {\bf D31} (1985) 1324.
   Later developments include
M.   Visinescu, {\it Phys. Lett.} {\bf B339}, (1994);
J. W. van Holten, {\it Phys. Lett.} {\bf B342}, 47 (1995),
A. Comtet and P. A. Horv\'athy,
 {\it Phys. Lett.} {\bf B349}, 49 (1995), etc.
 
  
\reference
R.~Jackiw,
Ann.~Phys. (N.Y.) {\bf 201}, 83 (1990).

 \reference
 C. J. Parks, Nucl. Phys. {\bf B367}, 99 (1992);
 J.-G. Demers, Mod. Phys. Lett. {\bf 8}, 827 (1993);
 C. Duval and P. A. Horv\'athy, Tours Preprint N. 60/93 (1993) (unpublished).

\reference
E.~Witten,
Nucl.~Phys. {\bf B185}, 513 (1981);
P.~Salomonson and J.W.~Van Holten,
Nucl.~Phys. {\bf B169}, 509 (1982);
M.~De Crombrugghe and V.~Rittenberg, Ann.~Phys. (N.Y.)
{\bf 151}, 99 (1983).

\reference
M.~Leblanc, G.~Lozano and H.~Min,
Ann. Phys. (N.Y.) {\bf 219}, 328 (1992);
C.~Duval and P.A.~Horv\'athy,
in Ref. [12].
The bosonic galilean symmetry was pointed out by
R. Jackiw and S.-Y. Pi, Phys. Rev. {\bf D 42}, 3500 (1990).

\reference
R. J. Hughes, V. A. Kosteleck\'y and M. M. Nieto,
Phys. Rev. {\bf D34}, 1100 (1986).

\reference
R.~Jackiw,
Phys.~Rev. {\bf D29}, 2375 (1984).
 
\reference
C. R. Hagen, Phys. Rev. Lett. {\bf 64}, 503 (1990);
R.~Musto, L.~O'Raifeartaigh and A.~Wipf,
Phys. Lett. {\bf B 175}, 433 (1986);
P. Forg\'acs, L.~O'Raifeartaigh and A.~Wipf,
Nucl. Phys. {\bf B 293}, 559 (1987).

\vfill\eject
\bye